\documentclass[graybox]{svmult}

\usepackage{helvet}         % selects Helvetica as sans-serif font
\usepackage{courier}        % selects Courier as typewriter font
\usepackage{type1cm}        % activate if the above 3 fonts are
\usepackage{makeidx}         % allows index generation
\usepackage{graphicx}        % standard LaTeX graphics tool
\usepackage{multicol}        % used for the two-column index
\usepackage[bottom]{footmisc}% places footnotes at page bottom
\usepackage{amsmath,amsfonts,amssymb,latexsym}
\usepackage{epsfig}

\def\ds{\displaystyle}
\def\be{\begin{equation}}
\def\ee{\end{equation}}

\def\bea{\begin{eqnarray}}
\def\eea{\end{eqnarray}}
\def\bean{\begin{eqnarray*}}
\def\eean{\end{eqnarray*}}
\def\H{{\mathcal H}}

\def\N{{\mathbb N}}
\def\R{{\mathbb R}}
\def\C{{\mathbb C}}

\def\S{{\mathbb S}}
\def\Z{{\mathcal Z}}
\def\a{{\alpha}}

\def\r{\rangle}
\begin{document}
\title*{$SU(2)$, Associated Laguerre Polynomials and Rigged Hilbert Spaces}

\author{Enrico Celeghini, Manuel Gadella and Mariano A del Olmo}

\institute{E. Celeghini \at Dpto di Fisica, Universit\`a  di Firenze and
INFN--Sezione di
Firenze, Firenze, Italy,
\at Dpto. de F\'{\i}sica Te\'orica, Universidad de
Valladolid,
E-47005, Valladolid, Spain\at\email{celeghini@fi.infn.it}
\and 
M. Gadella and M.A. del Olmo\at Dpto de F\'{\i}sica Te\'orica and IMUVA, Univ. de
Valladolid,
E-47005, Valladolid, Spain \at \email{manuelgadella1@gmail.com}\,,
\hskip0.15cm\email{marianoantonio.olmo@uva.es}
\and
Proceedings  of the 10-th International Symposium "Quantum Theory and Symmetries" (QTS10), 19-25 June 2017, Varna, Bulgaria (to be published  by  Springer).}
\maketitle

\abstract{We present a family of unitary irreducible representations of $SU(2)$ realized in the plane,
in terms of the Laguerre polynomials. These functions are similar to the spherical harmonics defined
on the sphere. Relations with an space of square integrable functions defined on the
plane, $L^2(\R^2)$, are analyzed. We have also enlarged this study using   rigged Hilbert spaces that allow  to work with 
 iscrete and continuous bases like is the case here.}

%%%%%%%%%%%%%%%%%%%%%%%%%%%%%%%%%%%%%%%%%%%%%%%%
%%%%%%%%%%%%%%%%%%%%%%%%%%%%%%%%%%%%%%%%%%%%%%%%
\section{Introduction}
\label{sec:1}

The representations of a Lie algebra are usually considered as ancillary to the algebra
and developed starting from  the algebra, i.e. from the generators and their commutation
relations.
The universal enveloping algebra (UEA) is constructed and
a complete set of commuting observables selected,  choosing between the invariant operators
of the algebra and of a chain of its subalgebras.
The common eigenvectors of this complete set of operators are a basis of a vector space where
the Lie algebra generators 
are realized as operators.

We propose here an alternative construction that allows to add to the representations
obtained following the reported recipe, new ones not achievable following the
previous approach. Starting from a concrete vector space of functions with discrete labels and continuous
variables, we consider the recurrence relations that allow to connect functions with
different values of the labels.  These recurrence relations are not operators but
allow us to introduce, for each label and for each continuous variable, an operator that reads its value.
In this way,  recurrence relations are rewritten in terms of rising and lowering operators built by means of
the above defined operators.These rising and lowering operators are often genuine generators of the Lie algebras considered
by Miller \cite{miller1977} and the procedure gives simply the representations of the algebras
in a well defined function space \cite{celeghini2013a,  celeghini2014}.
However it can happen that the commutators, besides the values required by the algebra, have
additional contributions. The essential point of this paper is that these additional contributions
(as exhibited here) can be proportional to the null identity that defines the starting
vector space.
As this identity is zero on the whole representation,
the Lie algebra is well defined and a new representation in a space of functions has been found.

We do not discuss here the general approach, but we
limit ourselves to a simple example where  all aspects are better understandable.
We start thus from the associated Laguerre functions (ALF) and, following the proposed construction,
we realize the algebra $su(2)$ in terms of the appropriate rising and lowering operators.
The ALF support in reality a larger algebra \cite{celeghini2015} but we prefer to consider
here only the subalgebra $su(2)$.
The reasons for this choice are twofold: first in this way the technicalities
are reduced at the minimum and second it has been very nice for us
to discover that not all representations of a so elementary group like $SU(2)$ where known.

As discussed in \cite{ceolgant, cedice16,celeghini2018} the presence of operators with spectrum of different
cardinality implies that, as considered for the first time in Lie algebras in
\cite{LiNa}, the space of the group representation is not a Hilbert space but a rigged
Hilbert space (RHS) \cite{bo}. Thus, 
 we introduce  the above setting within the context of RHS since the  RHS is the perfect framework where discrete and continuous bases coexist. In addition, the same RHS serves as a support for a representation on it of a Lie algebra as continuous operators  as well as  for its UEA. Therefore, the connection between discrete and continuous bases and Lie algebras with RHS is well established.

%%%%%%%%%%%%%%%%%%%%%%%%%%%%%%%%%%%%%%%%%%%%%%%%
%%%%%%%%%%%%%%%%%%%%%%%%%%%%%%%%%%%%%%%%%%%%%%%%

\section{Associated Laguerre polynomials}
\label{sec:2}

The ALP \cite{szego}, $L_n^{(\alpha)}(x)$,  
depend from a
real continuous variable $x \in [0 ,  \infty )$
and from two other real labels $(n,\a)$ : $n=0,1,2,\dots$  and $\alpha$
(usually assumed  as a fixed parameter) continuous and\, $>  -1$ . They reduce to the
Laguerre polynomials for\, $\alpha = 0$\,
and are defined by the second order differential equation
\be\label{defLag}
\left[x \frac{d^2}{dx^2} + (1+\alpha-x) \frac{d}{dx} + n \right] L^{(\alpha)}_n(x) =0 \;.
\ee
From the many recurrence relations that can be found in literature \cite{szego, NIST},
we consider the following ones, all first order differential recurrence relations:
\be
\begin{array}{l}\label{Moperators}
\displaystyle
\left[x \frac{d}{dx} + (n + 1 + a - x) \right] L_n^{(\alpha)}(x) =  (n + 1) L_{n+1}^{(\alpha)}(x)\,,
\\[0.35cm]
\displaystyle
\left[- x \frac{d}{dx} + n \right] L_n^{(\alpha)}(x) =  (n + \alpha) L_{n-1}^{(\alpha)}(x)\,, \quad
\\[0.35cm]
\displaystyle
\left[-\frac{d}{dx}  + 1 \right] L^{(\alpha)}_n(x) = L^{(\alpha+1)}_n(x)\,,
\\[0.35cm]
\displaystyle
\left[x \frac{d}{dx} + \alpha \right] L^{(\alpha)}_n(x) = (n + \alpha ) L^{(\alpha -1)}_n(x) \,.
\end{array}\ee
Starting from $L_{n}^{(\alpha)}(x)$, by means of repeated applications of
eqs.~(\ref{Moperators}), 
$L_{n+k}^{(\alpha+h)}(x)$ --with $h$ and $k$ arbitrary integers-- can be obtained
through a  differential relation of higher order.
But, by means of eq.~(\ref{defLag}), every differential relation of order two or higher
can be rewritten as a differential relation of order one. 
In particular we can obtain
\be \begin{array}{l}\label{recurrence2}
\displaystyle
\left[ \frac{d}{dx} + \frac{n}{\a+1} \right]\,L_n^{(\a)}(x) \,=\, -\, \frac{\a}{\a+1}\;
L_{n-1}^{(\a+2)}(x) ,\quad
\\[0.3cm]
\displaystyle
\left[ x (\alpha-1) \frac{d}{dx} -x\left(n+3\;\frac{\alpha}{2}\right)+\alpha(\alpha-1)\right]
\,L_n^{(\a)}(x) \\[0.25cm]
\hskip3.33cm\displaystyle = (j+\alpha)(\alpha+1) \; L_{n+1}^{(\a-2)}(x) ,
\end{array}\ee
that are the recurrence relations we employ in  this paper.

The ALP  $L_n^{(\alpha)}(x)$ are  --for $\alpha >-1$ and
fixed--  orthogonal
in $n$ with respect the weight measure $d\mu(x)=x^{\a}\, e^{-x}\, dx$ \cite{szego}:
\be \begin{array}{l}\label{orthogonality}
\displaystyle
\int_0^\infty\; dx \;x^\a\; e^{-x}\, L_n^{(\alpha)}(x)\; L_{n'}^{(\alpha)}(x)\; =\;
\frac{\Gamma(n+\a+1)}{n!}\;  \delta_{n n'}\; ,
\\ [0.3cm]
\displaystyle\
\sum_{n=0}^{\infty} \; x^\alpha \, e^{-x} \,L_n^{(\alpha)}(x)  L_n^{(\alpha)}(x')  =  \delta(x-x')\, .
\end{array}\ee

The parameter $\alpha$\, can be extended
to arbitrary complex values  \cite{szego} and, in particular, for $\alpha$ integer and such
that\; $0 \le |\alpha | \le n$\,, \,we have
 the relation
\be\label{aa}
L_n^{(-\alpha)}(x) = (-x)^\alpha \frac{(n-\alpha)!}{n!} \, L_{n-\alpha}^{(\alpha)}(x)\, .
\ee

Here we assume consistently that
$n \in \N \,,\; \alpha \in {\mathbb Z} $ and $ n-\alpha \in \N\,,$
and we also  consider\, $\alpha$\, as a label, like $n$, and not a parameter
fixed at the beginning.
Following the approach of \cite{celeghini2013a}, we introduce
now a set of alternative variables
and include the weight measure inside the functions, in such a way to obtain the
bases we are used in quantum mechanics.
We define indeed
$\ds  j := n + {\alpha}/{2} \,$ and $\ds m := -{\alpha}/{2} $  
that  are such that
$
j \in \N/2 \,,\;  j-m \in \N$ and $  |m| \le j\,.
$
Note that they look like the parameters $j$ and $m$  used in $SU(2)$.
Now we write
\[
{\cal L}_j^m(x) :=
\sqrt{\frac{(j+m)!}{(j-m)!}}\;
\, x^{-m} \, e^{-x/2} \,L_{j+m}^{(-2 m)}(x)
\]
so that, from eq.~(\ref{aa}), ${\cal L}_j^m(x)$ is symmetric/antisymmetric in the exchange
$m \leftrightarrow -m$ since
${\cal L}_j^m(x)=  (-1)^{2j} \,{\cal L}_j^{-m}(x)$.
From eqs.~(\ref{orthogonality}), we see that the ${\cal L}_j^m(x)$  verify, for $m$ fixed, the following orthonormality and completeness relations
 \be\label{orthnorm}
 \displaystyle \int_{0}^{\infty} {\cal L}^m_{j}(x)\, \, {\cal L}^m_{j'}(x) \;dx\;=
 \; \delta_{j j'}\,,\qquad
\displaystyle \sum_{j=|m|}^{\infty}  \; {\cal L}^m_{j}(x)\,\, {\cal L}^m_{j}(x')\; =
\; \delta(x-x')\,,
 \ee
and are thus, for any fixed value of $m$, an orthonormal basis of 
$L^2(\R^+)$.

Note that, in the algebraic description of the spherical harmonics,
the functions\,  $T_j^m(x)=
\sqrt{\frac{(j-m)!}{(j+m)!}}\, P_j^m(x) $, related to the associated Legendre functions $P_l^m(x)$
 and introduced in \cite{celeghini2013a},  satisfy $T_j^m(x) = (-1)^m\,T_j^{-m}(x)$ 
 which is a relation similar to those verified by the ${\cal L}_j^m(x)\,.$   Moreover the $T_j^m(x)$, like the ${\cal L}_j^m(x)$ on the half-line, are orthogonal --for fixed $m$-- in the
interval\; $(-1,+1)\subset \R$ and a basis for $L^2[-1,1]$.

%%%%%%%%%%%%%%%%%%%%%%%%%%%%%%%%%%%%%%%%
%%%%%%%%%%%%%%%%%%%%%%%%%%%%%%%%%%%%%%%%
\section{$SU(2)$  representations in the plane}

Following now Ref.~\cite{celeghini2013a}, we define four  operators
$X$, $D_x$, $J$ and $M$ such that
\be\begin{array}{ll}\label{fouroperators}
\displaystyle
X\; {\cal L}^m_j(x)\; &=\; x\; {\cal L}^m_j(x),\;\quad\qquad D_x\; {\cal L}^m_j(x)\;
=\; {\cal L}^m_j(x)' \;,
\\[0.3cm]
\displaystyle
J\; {\cal L}^m_j(x)\; &=\; j\; {\cal L}^m_j(x),\;\quad\qquad M\; {\cal L}^m_j(x)\;
=\; m\; {\cal L}^m_j(x) .
\end{array}\ee
and we can rewrite eq.~(\ref{defLag}) in terms of the ${\cal L}^m_j(x)$ and in operatorial form as
\be\label{XXX}
E\, {\cal L}^m_j(x)\, \equiv\,\left[ X\, D_x^2 + D_x
-\frac{1}{X} M^2-\frac{X}{4} +J+\frac{1}{2}\,\right]\,{\cal L}^m_j(x)=0 \,.
\ee
Thus, the identity\, $E \equiv 0$ \,
defines  
$L^2(\R^+)$.

The relations (\ref{recurrence2}) can now be rewritten on terms of the\, ${\cal L}_j^m(x)$  as
\be\label{actionKpm}
\begin{array}{lll}
K_+\,{\cal L}_j^m(x)&=&\ds \sqrt{(j-m)(j+m+1)}\;\; {\cal L}_j^{m+1}(x)\,,
\\[0.3cm]
K_-\,{\cal L}_j^m(x)&=&\ds \sqrt{(j+m)(j-m+1)}\;\; {\cal L}_j^{m-1}(x) \,,
\end{array}\ee
where
\be\label{Kpm}
\begin{array}{lll}
K_+&=&\ds - 2 D_x \left(M+\frac{1}{2}\right)+\frac{2}{X} M\left(M+\frac{1}{2}\right) -
\left(J+\frac{1}{2}\right) \,,
\\[0.6cm]
K_-&=&\ds  2  D_x \left(M-\frac{1}{2}\right)+\frac 2 X M\left(M-\frac{1}{2}\right) -
\left(J+\frac{1}{2}\right)\, .
\end{array}\ee
Since, from eqs.~(\ref{actionKpm}), we have
$
[K_+,K_-]\;\,  {\cal L}_j^m(x)\,=\; 2\, m \;{\cal L}_j^m(x)
$
and assuming   $K_3\, :=\, M$  (i.e. 
$
K_3\;\,  {\cal L}_j^m(x)= \, m \,\;{\cal L}_j^m(x) 
$)
we get the relations
\be\label{conmutadoresK}
[K_+,K_-]\;\,  {\cal L}_j^m(x)\;
 =\; 2\, K_3\;\, {\cal L}_j^m(x) ,
\;\;
[K_3, K_\pm]\;  {\cal L}_j^m(x)\;
 =\;  \pm\, K_\pm\;{\cal L}_j^m(x)\,,
\ee
that display the fact that, for fixed\, $j$, under the action of  $K_\pm$ and $K_3$\,, the   
${\cal L}_j^m(x)$
supports  the irreducible representation of dimension $2 j+1$ of $su(2)$.

However, while as exhibited by (\ref{orthnorm}) the space $\{{\cal L}_j^m(x)\}$ has an inner product for $m$ fixed and $j \geq |m|$ 
(thus supporting a set of UIR of $SU(1,1)$ \cite{celeghini2015}),
the representation \eqref{conmutadoresK} of $SU(2)$ is not faithful, since ${\cal L}_j^m(x)= (-1)^{2 j}\,  {\cal L}_j^{-m}(x)$,  and not unitary.
The definition of a scalar product is indeed one of the problems we have in the connection of hypergeometric functions and Lie algebras. Hence,
we have two problems: the ${\cal L}_j^m(x)$ are not orthonormal
for $j$ fixed and  functions with opposite $m$ are not independent (as it happens also with the $P_j^m(x)$). Following the same approach of the spherical harmonics to construct  the inner product space  for $j$ fixed and $|m|\leq j$ we, thus, introduce  a  new real variable  $\phi$  ($ -\pi <\phi \le \pi $) and the new objects
\[
{\cal Z}_j^m(r,\phi) :=  e^{i\, m\, \phi}\, {\cal L}_j^m(r^2),
\]
  that verify
  $
  {\cal Z}_j^m(r,\phi+2\pi) =(-1)^{2 j}\,{\cal Z}_j^m(r,\phi) .
  $
  Under the change of variable $x\to r^2$ equation \eqref{XXX}  becomes for ${\cal Z}_j^m(r,0)$
  \be
\label{ODEL2}
\left[\frac{d^2}{dr^2} + \frac{1}{r} \frac{d}{dr}
-\frac{4 m^2}{r}- r^2 + 4(j+\frac{1}{2})\right]\,\,{\cal Z}^m_j(r,0)
=0 .
\ee
The functions ${\cal Z}_j^m(r,\phi)$ are the analogous on the plane  of the
spherical harmonics ${Y_{lm}(\theta,\phi)}$ on the sphere.
The orthonormality and completeness of the
${\cal Z}_j^m(r,\phi)$  is similar to that of $Y_j^m(\theta,\phi)$
\be \begin{array}{l}\label{OrthNormZ}\displaystyle
\frac{1}{ \pi} \int_{-\pi}^{\pi} d \phi \int_{0}^{\infty} \; r\, dr\;\; {\cal Z}^m_{j}(r, \phi)^*\,
\, {\cal Z}^{m'}_{j'}(r, \phi)\; =\; \delta_{j,j'}\;\; \delta_{m,m'} \;,
\\[0.5cm]
\displaystyle \sum_{j, m}  \;\; {\cal Z}^m_{j}(r, \phi)^*\,\, {\cal Z}^m_{j}(r', \phi')\;
=\; \frac{\pi}{r}\;\delta(r-r')\; \delta(\phi-\phi') \,.
\end{array}\ee
This means that  $\{{\cal Z}_j^m(r, \phi)\}$ is a basis of the Hilbert space  $L^2(\R^2)$ with  measure 
$d\mu (r,\phi )=r\,dr\,d\phi/\pi$ like
$\{Y_j^m(\theta,\phi)\}$ is a basis of  $L^2(S^2)$ with $d\Omega$. 

Now we consider an abstract Hilbert space ${\cal H}$  supporting the $2j+1$ dimensional  IR of $su(2)$   spanned  by the eigenvectors of  $J$ and $M$  (see eq. \eqref{fouroperators}) 
\[
J\,|j,m\rangle=j\,|j,m\rangle\,,\quad
M\,|j,m\rangle=m\,|j,m\rangle\,,\qquad 2 j\in \N,\;\; |m|\leq j\,.
\]
These vectors $|j,m\rangle$ constitute a basis of ${\cal H}$ verifying the properties of orthogonality and completeness
\[
\langle j,m|j',m'\rangle=\delta_{j,j'}\,\delta_{m,m'}\,,\qquad
\sum_{j=0}^{\infty}\sum_{m=-j}^{j} |j,m\rangle\langle j,m|=I
\] 
Any $|f\r \in\mathcal H$ may be written as
$|f\r=\sum_{j=0}^\infty\sum_{m=-j}^j f_{j,m}\,|j,m\rangle$ 
if and only if 
\begin{equation}\label{14}
 \sum_{j=0}^\infty\sum_{m=-j}^j |f_{j,m}|^2<\infty\,,\qquad  f_{l,m}=\langle l,m|f\rangle\,.
\end{equation}

A canonical injection  $S:{\mathcal H}\to L^2(\R^2)$ can be defined by $|j,m\rangle\to {\cal Z}^m_{j}(r, \phi)$
 and extended by linearity and continuity to the whole $\mathcal H$. One can easily  check that $ S$ is unitary.  For any 
 $|f\r \in\H$  we have the following expression
\[\label{16}
S|f\r =\sum_{j=0}^\infty\sum_{m=-j}^j f_{j,m}\,S\,|j,m\rangle
=\sum_{j=0}^\infty\sum_{m=-j}^j f_{j,m}\,{\cal Z}^m_{j}(r, \phi)\,.
\]

We  now introduce a continuous basis, $\{|r,\phi\rangle\}$, depending on the values of the  variables  $r$ and  $\phi$ with the help of the discrete basis $\{|j,m\r\}$ by
\begin{equation}\label{17}
\langle r,\phi |j,m\rangle :=\Z_j^m(r,\phi)\,.
\end{equation}
In reality, because of the different cardinality of $r$ and $j$, we are dealing with   a RHS (see next Section).  
The $
{\cal Z}_j^m(r,\phi)$ can be seen as 
  the transformation matrices from the irreducible representation states $\{|j,m\rangle\}$  to the localized states in the plane $\{|r,\phi\rangle\}$, like
$
Y_j^m(\theta,\phi)=\langle j,m|\theta,\phi\rangle
$
 are the corresponding ones to the localized states  $\{|\theta,\phi\rangle\}$  in the sphere \cite{celeghini2018,wu}. Indeed
\[
|j,m\rangle =\frac{1}{\pi}\int_{\R^2} |r, \phi\rangle  {\cal Z}_j^m(r,\phi)   r dr d\phi ,\quad
|j,m\rangle = \int_{S^2} |\theta, \phi\rangle \sqrt{j+1/2}Y_j^m(\theta,\phi)  d\Omega .
\]

We continue with the analogy and, from  $K_\pm$ and $ K_3 $ (\ref{Kpm}), we  define
\be\label{alg}
{J}_\pm \,:=\, e^{\pm i \phi}\, K_\pm, \qquad\;\; J_3 \,:=\, K_3,
\ee
with  act on the ${\Z}_j^m(r,\phi)$ as
\be\label{actionJ}
\begin{array}{lll}
J_+\;\; {\Z}_j^m(r,\phi)&=&\ds \sqrt{(j-m)(j+m+1)} \;\, {\Z}_j^{m+1}(r,\phi),\\[0.4cm]
J_-\;\;  {\Z}_j^m(r,\phi)&=&\ds \sqrt{(j+m)(j-m+1)} \;\, {\Z}_j^{m-1}(r,\phi),\\[0.4cm]
J_3\;\; {\Z}_j^m(r,\phi)&=&\ds m \; {\Z}_j^m(r,\phi) \,.
\end{array}\ee 
The functions ${\cal Z}_j^m(r,\phi)$ with $j$ fixed and $|m|\leq j$, are
orthonormal and determine the representation of
dimension $2 j+1$ of $su(2)$ as it happens  for the  $Y_j^m(\theta, \phi)$.
However there is a essential difference between the operators  $\{ J_\pm, J_3\}$
that act on the sphere $\S^2$  that are true  generators of $su(2)$ and the
 $\{ J_\pm, J_3\}$ of (\ref{alg}), defined in $\R^2$, that do not
close a Lie algebra.
Indeed, when we calculate the commutator $[ J_+, J_- ]$ in terms of the differential operators defined in
the eqs.~\eqref{Kpm} and (\ref{alg}), we obtain
$\ds
[ J_+, J_- ]\; =\; 2\, J_3 + \frac{8}{R^2}\, J_3 \,E\,,
$
 and  only when
$E \equiv 0$\,,\, i.e. only in the unitary space \, $L^2(\R^2)$ ,
the $su(2)$ algebra is recovered.
On the other hand, $E$ is related to the $su(2)$ Casimir  $\cal C$
\[\label{cascon}
E\,=\, -\frac{R^2}{4 J_3^2+1}\,\left[ {\cal C}- J(J+1)\right]\,\equiv \, -\frac{R^2}{4 J_3^2+1}\,\left[ J_3^2 +\frac{1}{2}\; \{ J_+, J_- \}\;  - J(J+1)\right],
\]
 so equation $E=0$ is equivalent to the  $su(2)$ Casimir condition ${\cal C}- J(J+1)=0$, that entails the usual Lie algebra in each $su(2)$ representation space.

%%%%%%%%%%%%%%%%%%%%%%%%%%%%%%%%%%%%%%%%%%%
%%%%%%%%%%%%%%%%%%%%%%%%%%%%%%%%%%%%%%%%%%%%
\section{Rigged Hilbert space formulation.}

A RHS  (or Gelf'and triplet)
 is a triplet of spaces  
$
\Phi\subset\mathcal H\subset \Phi^\times\, ,
$
where
$\mathcal H$   is an infinite dimensional separable Hilbert space, $\Phi$  is   a dense subspace of $\mathcal H$ endowed with its own topology, and
$\Phi^\times$ is the dual (or the antidual) space of $\Phi$ \cite{bo,GEL,BG}. 
The  topology considered on $\Phi$ is finer  (contains more open sets) than the topology that 
$\Phi$ has as subspace of $\mathcal H$, and
$\Phi^\times$ is equipped with a topology compatible with the dual pair
 $(\Phi,\Phi^\times)$ \cite{HOR}, usually the weak topology.
The topology of $\Phi$ \cite{PI,GG1} allows that all sequences which converge on $\Phi$, also converge on $\mathcal H$ but the converse is not true. 
The difference between topologies gives rise that   $\Phi^\times$ is bigger than $\mathcal H$, which is self-dual. 

Here, 
 any $F\in\Phi^\times$ is a continuous linear mapping from $\Phi$ into $\C$.

An essential property   is that 
if   $A$ is a densely defined operator on $\mathcal H$, such that $\Phi$ be a subspace of its domain and that 
$ A\varphi\in\Phi$ for all $\varphi\in\Phi$,
  we say that $\Phi$ reduces $A$ or that $\Phi$ is invariant under the action of $A$, (i.e., $A\Phi\subset\Phi$).  
Then $A$ may be extended unambiguously to  $\Phi^\times$ by   the duality formula
\be\label{dualidad}
\langle A^\times\,F|\varphi\rangle :=\langle F|A\varphi\rangle\,,\qquad \forall\,\varphi\in\Phi\,, \;\;\forall\,F\in\Phi^\times\,.
\ee
Moreover if $A$ is continuous on $\Phi$, then $A^\times$  is continuous on $\Phi^\times$.

The  topology on $\Phi$ is given by an infinite countable set of norms
$\{||-||_{n=1}^\infty\}$.
A linear operator $A$ on $\Phi$ is continuous  if and only if for each norm $||-||_n$ there is a  $K_n>0$ and a finite sequence of norms $||-||_{p_1}, ||-||_{p_2}, \dots, ||-||_{p_r}$ such that for any $\varphi\in\Phi$, one has \cite{RS}
\begin{equation}\label{continuidad}
||A\varphi||_n\le K_n\,\left(||\varphi||_{p_1}+||\varphi||_{p_2}+\dots+||\varphi||_{p_r}\right)\,.
\end{equation}

Now let us go to define and use the RHS $\mathfrak G\subset\mathcal H\subset\mathfrak G^\times$ where discrete and continuous bases coexist and the meaningful operators are well defined and continuous. Since we have a representation in terms of the ${\cal Z}_j^m(r,\phi)$, it would be more convenient to start with an equivalent RHS
$
\mathfrak D\subset L^2(\R^2)\,\subset\,\mathfrak D^\times\,,
$ 
such as  $\mathfrak D$ is a test functions space with  $f(r,\phi)\in L^2(\R^2)$, which therefore admit the span 
\begin{equation}\label{7}
f(r,\phi)=\sum_{j=0}^\infty \sum_{m=-j}^j  f_{j,m}\,{\Z}_j^m(r,\phi)\,,
\end{equation}
where the series converges in the sense of the norm in $L^2(\R^2)$. A  necessary and sufficient condition for it is 
$
\sum_{j=0}^\infty \sum_{m=-j}^j | f_{j,m}|^2<\infty\,.
$ 
Thus, from \eqref{7}, we define  $\mathfrak D$ as the space of functions $f(r,\phi)$ in $L^2(\R^2)$ such that
\begin{equation}\label{40}
||f(r,\phi)||_n^2:= \sum_{j=0}^\infty  \sum_{m=-j}^j\,(j+|m|+1)^{2n}\, |f_{j,m}|^2<\infty\,,\qquad n=0,1,2,\dots\,.
\end{equation}
Obviously, all the finite linear combinations of the  ${\Z}_j^m(r,\phi)$ are in $\mathfrak D$, hence $\mathfrak D$ is dense in $L^2(\R^2)$. Thus, the family of norms $||-||_n$ on $\mathfrak D$ \eqref{40} gives a topology such that $\mathfrak D$ is a Fr\`echet space (metrizable and complete). Since for $n=0$ we have the Hilbert space norm, the canonical injection from $\mathfrak D$ into $L^2(\R^2)$ is continuous.

Because $j$ goes from $0$ to $\infty$, the operators $J_\pm,J_3$ are all unbounded and, therefore, their respective domains are densely defined on  $L^2(\R^2)$, but not on the whole $L^2(\R^2)$. 
We can  prove that all these operators are defined on the whole $\mathfrak D$ and are continuous with the topology on $\mathfrak D$. The proof is simple and it is essentially the same for all  operators. As an example, let us give the proof for $J_+$. For any function $f$ in $\mathfrak D$, we have $J_+ f$, i.e., 
\begin{eqnarray*}\label{42}
J_+ \sum_{j=0}^\infty  \sum_{m=-j}^j f_{j,m}\,{\cal Z}_j^m(r,\phi)= \sum_{j=0}^\infty  \sum_{m=-j}^j f_{j,m}\,\sqrt{(j-m)(j+m+1)}\,{\cal Z}_{j}^{m+1}(r,\phi)\,.
\end{eqnarray*}
To show that $J_+ f\in\mathfrak D$ we have to prove that for any $n\in\N$, it satisfies (\ref{40}). So taking into account the shift on the index $m$ \eqref{actionJ} we have
\begin{eqnarray}\label{43}
\sum_{j=0}^\infty  \sum_{m=-j}^j  |f_{j,m}|^2\,(j-m)(j+m+1)\,(j+1+|m|+1)^{2n}\,.
\end{eqnarray}
The following two inequalities are straightforward:
\begin{equation*}\label{444}
(j-m)(j+m+1)\le (j+|m|+1)^2\,,\qquad
(j+1+|m|+1)^{2n}\le 2^{2n}\,(j+|m|+1)^{2n}\,.
\end{equation*}
Using these inequalities   we see that \eqref{43} is bounded by 
\begin{eqnarray}\label{45}
2^{2n} \sum_{j=0}^\infty  \sum_{m=-j}^j  |f_{j,m}|^2\,(j+1+|m|+1)^{2n+2}\,,
\end{eqnarray}
which  converges after (\ref{40}). Hence,  $J_+\,f\in \mathfrak D$. In order to show the continuity of $J_+$ on $\mathfrak D$, we use  (\ref{continuidad}). Thus,  applying $J_+$ to  any $f(r,\phi)\in \mathfrak D$   we get
\begin{equation*}\label{46}
||J_+f(r,\phi)||_n^2\le 2^{2n}\,||f(r,\phi)||_{n+1}^2 \Longrightarrow
  ||J_+f(r,\phi)||_n \le 2^n\,||f(r,\phi)||_{n+1}\,,
\end{equation*}
which satisfies (\ref{continuidad}) for all $n=0,1,2,\dots$.  Hence, the continuity of $J_+$ on $\mathfrak D$ has been proved. By means of the duality formula, we extend $J_+$ to a weakly continuous operator on $\mathfrak D^\times$. Same properties can be proved for $J_-$ and $J_3$. 

Now we are able  to define the  abstract RHS $\mathfrak G\subset\mathcal H\subset \mathfrak G^\times$ using the unitary mapping $S:\H\to L^2(\R^2)$  introduced in the previous section. Thus, we define $\mathfrak G:=S^{-1}\mathfrak D$. Hence the topology on $\mathfrak G$ is the transported topology from $\mathfrak D$ by $S$, so that if $f\in\mathfrak G$, the semi-norms are
\begin{equation*}\label{47}
||f||_n^2= \sum_{j=0}^\infty  \sum_{m=-j}^j\,(j+|m|+1)^{2n}\, |f_{j,m}|^2<\infty\,,\qquad n=0,1,2,\dots\,.
\end{equation*}
The topology on $\mathfrak G$ uniquely defines $\mathfrak G^\times$. Moreover there exists a one-to-one continuous mapping from $\mathfrak G$ onto $\mathfrak D$ with continuous inverse. It is given by an extension, $\widetilde S$, of $S$ defined via the duality formula $\langle \widetilde S f |\widetilde S F\rangle=\langle f|F\rangle$, with $f\in\mathfrak G$ and $F\in\mathfrak G^\times$. 

On the other hand, if an operator $O$ satisfies $O\mathfrak D\subset \mathfrak D$ with continuity, the same property works for $\widehat O=S^{-1}OS$ on $\mathfrak G$.

%%%%%%%%%%%%%%%%%%%%%%%%%%%%%%%%%%%%%%%%%%%
%%%%%%%%%%%%%%%%%%%%%%%%%%%%%%%%%%%%%%%%%%%
\section{Conclusions}
Starting from the recurrence relations (\ref{recurrence2}) we   obtained
the operators $\{ J_\pm, J_3 \}$ (\ref{alg}). Their general linear algebra is not a Lie algebra.
However its representation  on  $L^2(\R^2)$,
characterized by the eigenvalue zero of the operator $E$,
is isomorphic to the regular representation $\{|j,m\rangle\}$ of $su(2)$ and it has therefore  a stronger symmetry than the  general linear operator structure itself.

We are used in Lie algebra theory to representations that preserve the symmetry of the algebra and
to algebras that have the same symmetry of the
space where the representation is defined.
This is exactly what happens with the spherical harmonics, that are solution of Laplace equation and,
thus, have the same intrinsic symmetry of the group\, $SU(2)$ of which they are representation bases.
However, here the situation is different since we represent $SU(2)$ in the plane $\R^2$ which geometry
preserves only the subgroup $SO(2)$ of $SU(2)$.
  Indeed
 $\{ J_\pm, J_3\}$  (\ref{alg}) are defined
for arbitrary $E$, but they
generate $su(2)$ only under the assumption $E\equiv0$, i.e. when we restrict ourselves to functions $f$ verifying the Casimir condition
$
{\cal C}\, f = J(J+1)\,f ,
$, i.e. that belong to $L^2(\R^2)$.

Reversing the connection, the  representations of a Lie algebra have been  related
not only to the Lie algebra itself but also
 to a set of operators that do not close a Lie algebra in an universal way but reduce to a Lie algebra only when   applied to well defined vector spaces.
 
 This paper offers a method to introduce representations of Lie groups in spaces
that are not symmetric under the group action and in situations where the general linear group of operators is
not a Lie group in a  universal way.

We have also constructed  two RHS ($\mathfrak G\subset\mathcal H\subset \mathfrak G^\times$ and $\mathfrak D\subset L^2(\R^2)\,\subset\,\mathfrak D^\times$)  supporting two UIR of $SU(2)$, the first one is related with the discrete basis 
$\{|j,m\r\}$ and the other RHS with the continuous one $\{|r,\phi\}$. Both are related by the unitary map $S: |j,m\r\to \Z_j^m(r,\phi)$ that also transports the topologies of the first RHS and other properties to the second RHS.
%%%%%%%%%%%%%%%%%%%%%%%%%%%%%%%%%%%%%%%%%%%%%%%%
%%%%%%%%%%%%%%%%%%%%%%%%%%%%%%%%%%%%%%%%%%%%%%%%

\begin{acknowledgement}
Partial financial support is acknowledged to the  Junta de Castilla y Le\'on and FEDER (Project VA057U16) and MINECO of Spain (Project MTM2014-57129-C2-1-P).
\end{acknowledgement}
%%%%%%%%%%%%%%%%%%%%%%%%%%%%%%%%%%%%%%%%%%%%%%%%
%%%%%%%%\section*{References}%%%%%%%%%%%%%%%%%%%%%%%%%%%%%

%%%%%%%%%%%%%%%%%%%%%%%%%%%%%%%%%%%%%%%%%%%%%%%%
%%%%%%%%%%%%%%%%%%%%%%%%%%%%%%%%%%%%%%%%%%%%%%%%

\end{document}